\documentstyle[preprint,aps,prd]{revtex}
\tightenlines
\newcommand{\be}{\begin{equation}}
\newcommand{\ee}{\end{equation}}
\newcommand{\ba}{\begin{eqnarray}}
\newcommand{\ea}{\end{eqnarray}}
\begin{document}

\title{Gravitomagnetic moments of the fundamental fields}
\author{R. Aldrovandi, V. C. de Andrade and J. G. Pereira}
\vskip 0.5cm
\address{Instituto de F\'{\i}sica Te\'orica \\
Universidade Estadual Paulista \\
Rua Pamplona 145 \\ 01405-900 S\~ao Paulo SP \\
Brazil}
\maketitle
\begin{abstract}
The quadratic form of the Dirac equation in a Riemann spacetime yields a
gravitational gyromagnetic ratio ${\kappa}_S = 2$
for the interaction of a Dirac spinor with curvature. A gravitational
gyromagnetic ratio ${\kappa}_S = 1$ is also found for the interaction of 
a vector field with curvature. It is
shown that the Dirac equation in a curved background can be obtained as the
square--root of the corresponding vector field equation only if the gravitational
gyromagnetic ratios are properly taken into account.
\end{abstract}

\vskip 2.0cm


\vskip 1.0cm


\vfill \eject

\section{Introduction}

The interaction of the spins of fundamental fields with gravitation
is a problem not yet fully understood. The correct value of the gravitomagnetic
moment associated to spin~\cite{wheeler}, for example, is a question
still asking for a consistent answer~\cite{hehl1}. Through the study of the
coupling of fundamental fields to curvature~\cite{khri}, and by analyzing the
self--consistency of the corresponding field equations, we undertake in this letter to
get a better understanding of those problems . The approach will be straightforwardly
field--theoretical. Use will be made of the tetrad formalism~\cite{livro} which, in
contrast to the metric approach, is able to describe the gravitational coupling of
both integer and half--integer spin fields. The greek  alphabet
($\mu$,$\nu$,$\rho$,~$\cdots=1,2,3,4$) will be used to denote tensor indices,  
that is, indices related to spacetime. Latin letters ($a$, $b$,
$c$,~$\cdots = 1,2,3,4$) will indicate local Lorentz (or Minkowski tangent space)
indices. Tensor and local Lorentz indices can be changed into each other through
contractions with the tetrad, denoted by $h^{a}{}_{\mu}$, and which satisfy
\[
h^a{}_{\mu} \, h_a{}^\nu = \delta_{\mu}{}^{\nu} \quad ; \quad
h^a{}_{\mu} \, h_b{}^\mu = \delta^{a}{}_{b} \; .
\]

The interaction of an arbitrary field with gravitation is usually taken into
account via a minimal coupling prescription, according to which the Minkowski metric
is replaced by the corresponding riemannian metric,
\begin{equation}
\eta_{a b} \rightarrow g_{\mu \nu} = \eta_{a b} \, h^{a}{}_{\mu}
h^{b}{}_{\nu} \; ,
\label{gmn0}
\end{equation}
and all ordinary derivatives, once applied to tensor objects, are replaced by
covariant derivatives,
\begin{equation}
\partial_c \rightarrow {\stackrel{\circ}{\nabla}}{}_{\mu} =
\partial_\mu - {\stackrel{\circ}{\Gamma}}{}_{\mu} \; ,
\end{equation}
with ${\stackrel{\circ}{\Gamma}}{}_{\mu}$ the Levi--Civita connection of the metric
$g_{\mu \nu}$. The General Relativity approach, thus, accounts only for tensor fields.
To include also spinor fields, a more comprehensive minimal coupling prescription must
be used. According to this generalized prescription, all ordinary derivatives must be
replaced by Fock--Ivanenko derivative operators~\cite{fockivan}
\begin{equation}
\partial_c \rightarrow {\mathcal D}_\mu = \partial_\mu - \frac{i}{2} \,
{\stackrel{\circ}{\omega}}{}^ {a b}{}_{\mu} \, J_{a b} \; ,
\label{mcp0}
\end{equation}
where~\cite{dirac}
\begin{equation}
{\stackrel{\circ}{\omega}}{}^{a}{}_{b \mu} = h^{a}{}_{\rho} \,
{\stackrel{\circ}{\nabla}}{}_{\mu} \, h_{b}{}^{\rho} \equiv h^{a}{}_{\rho}
\left( \partial_\mu h_b{}^\rho + {\stackrel{\circ}
{\Gamma}}{}^{\rho}{}_{\nu \mu} h_b{}^\nu \right)
\label{dirac}
\end{equation}
is the spin connection and $J_{a b}$ is, for each field under consideration,
the generator of the appropriate representation of the Lorentz group acting
on the Minkowski tangent space~\cite{wigner}. The explicit form of the
Fock--Ivanenko derivative, therefore, depends on the spin character of the
field as defined by the Lorentz transformations {\it on the tangent space}. 

Tensor indices, as said, are transcribed from spacetime to tangent  space, and
vice-versa, by simple contractions with the tetrad. To any real {\it Lorentz
tensor} $A^{a \dots b}$, that is, a tensor field transforming according to an
integer--spin representation of the Lorentz group, there is a corresponding
{\it spacetime tensor}
$
A^{\rho \dots \sigma} =  h_{a}{}^{\rho} \dots h_{b}{}^{\sigma} \,
A^{a \dots b} \; ,
$
that is, a field transforming as a tensor under a general spacetime coordinate
transformation. Consequently, to any Fock--Ivanenko derivative of a
Lorentz tensor will correspond a covariant derivative of a spacetime tensor. In
effect, by using the generators of the appropriate tensor representation of the Lorentz
group, the Fock--Ivanenko derivative of an integer--spin field $A^{a \dots b}$ can
always be written as a Levi--Civita covariant derivative of the corresponding
spacetime tensor $A^{\rho \dots \sigma}$:
\begin{equation}
{\mathcal D}_\mu A^{a \dots b} = h^{a}{}_{\rho} \dots h^{b}{}_{\sigma} \, 
{\stackrel{\circ}{\nabla}}{}_{\mu} A^{\rho \dots \sigma} \; .
\label{fivec4}
\end{equation}
For these fields, therefore, the minimal coupling prescription (\ref{mcp0})
can be restated as
\begin{equation}
{\partial}_c A^{a \dots b} \rightarrow 
{\stackrel{\circ}{\nabla}}{}_{\mu} A^{\rho \dots \sigma} \; ,
\label{mcp1}
\end{equation}
which is the form usually found in the literature~\cite{birrel}. 

Things are different for half--integer spin fields. There exists no spinor
representation of the group of general coordinate transformations, under which a
spinor field behaves actually as a scalar~\cite{velt}. In consequence, there is no a
Levi--Civita covariant derivative of a spinor field. The minimal
coupling prescription for these fields, therefore, must remain that given by
Eq.~(\ref{mcp0}), taking into account the spin character as defined {\it on the tangent
space}. This implies the simultaneous  presence of two different kinds of indices in a
theory describing the interaction of spinors with gravitation: tangent Lorentz and
spacetime tensor indices. This, in turn, leads to the explicit presence of the
tetrad field connecting these two kinds of indices. This is the reason why the
interaction of spinor fields with gravitation, as is usually said, requires a
tetrad formalism.

\section{Gravitomagnetic Moment}

The magnetic moment $\mbox{\boldmath$m$}_{L}$ of an electric
current distribution $\mbox{\boldmath$j(r)$}=\rho_e \mbox{\boldmath$v(r)$}$ is
usually defined as
\begin{equation}
\mbox{\boldmath$m$}_{L} = \frac{1}{2 c} \, \int \mbox{\boldmath$r \times j(r)$}
\, d^{3}r \; .
\label{magmo}
\end{equation}
For the specific case of an electron in a circular orbit,
\begin{equation}
\frac{m_{L}}{L} = \frac{e}{2 m c} \equiv \frac{g_{L} \; m_B}{\hbar} \; ,
\end{equation}
where $L$ is the orbital angular momentum of the electron, $g_{L} = 1$ is the
orbital gyromagnetic ratio, and
$m_B = e \hbar/2 m c$
stands for the Bohr magneton, the usual unit of magnetic moment. In an
analogous way, we can define the gravitomagnetic moment $\mbox{\boldmath$\mu$}_{L}$
of a mass current distribution $\mbox{\boldmath$p(r)$}=\rho_m
\mbox{\boldmath$v(r)$}$ as
\begin{equation}
\mbox{\boldmath$\mu$}_{L} = \frac{1}{2 c} \, \int \mbox{\boldmath$r \times p(r)$}
\, d^{3}r \; .
\label{gmagmo}
\end{equation}
If we consider again an electron in a circular orbit, we get
\begin{equation}
\frac{\mu_{L}}{L} = \frac{1}{2 c} \equiv \frac{\kappa_{L} \;
\mu_B}{\hbar} \; ,
\end{equation}
where $\kappa_{L}=1$ is the orbital gravito--gyromagnetic ratio, and
$\mu_B = \hbar/2 c$
is the gravitational analog of the Bohr magneton, the unit in which the
gravitomagnetic moment is to be measured. Notice that $\mu_B$ can be obtained
from $m_B$ by changing the {\it electric charge} $e$ by the {\it gravitational
charge} $m$. The absence of $m$ in the gravitational magneton is a beautiful 
manifestation of the universality of gravitation.

The electron intrinsic angular momentum, its spin $\mbox{\boldmath$S$}$, gives
rise to an intrinsic magnetic moment $\mbox{\boldmath$m$}_S$, usually written as
\begin{equation}
m_S = \frac{g_S \; m_B}{\hbar} \, S \; ,
\label{ms}
\end{equation}
with $g_S$ the spin gyromagnetic ratio. Again by analogy, we can say that the
electron intrinsic angular momentum gives rise also to a gravitomagnetic
moment
\begin{equation}
\mu_S = \frac{\kappa_S \; \mu_B}{\hbar} \, S \; ,
\label{mis}
\end{equation}
with $\kappa_S$ the spin gravito--gyromagnetic ratio.

\section{The Squared Dirac Equation}

The electromagnetic field is described by the potential $A_\mu$, which is a
connection on the corresponding gauge bundle. The value of the electron gyromagnetic
ratio
is obtained from the interaction between the intrinsic electron magnetic moment
(\ref{ms}) and the magnetic field. The latter is represented by the magnetic components
of the field strength $F_{\mu \nu}$, the curvature of the connection $A_\mu$.
Analogously, the value of the gravitational gyromagnetic ratio must be obtained
from the interaction between the intrinsic electron gravitomagnetic moment (\ref{mis})
and the gravitomagnetic components of the Riemann tensor, the field strength
of gravitation, which is given by the curvature of the Levi--Civita
connection.
We notice that this concept of gravitational gyromagnetic ratio~\cite{hehl2} is different
from the notion adopted by some authors, who consider the coupling to the so
called ``gravitomagnetic field"~\cite{wheeler}, a certain component of a weak--gravitational
field that is proportional to the Levi--Civita connection, not to the
curvature. In this context, a gravitational gyromagnetic ratio 1 has
been found~\cite{tiomno}, but this does not refer to the coupling with the Riemann
tensor, that is, to the field strength of the gravitational field, and is not,
therefore, the gravitational analog of the gyromagnetic ratio.

As is well known, the Dirac equation yields the value $g_S = 2$ for the
gyromagnetic ratio. This can be found either by taking the non-relativistic
limit yielding the Pauli equation, which shows its essentially non--relativistic
character, or by squaring the Dirac equation itself~\cite{itzy}. To find the
value of the gravitational gyromagnetic moment $\kappa_S$,
we shall follow a procedure analogous to the second approach. This means that we
shall take the Dirac equation in a Riemann spacetime~\cite{dirac},
\begin{equation}
i \hbar \, \gamma^\mu \, {\mathcal D}_\mu \psi - m c \psi = 0 \; ,
\label{diraceq}
\end{equation} 
with
$
\gamma^\mu = h_{a}{}^{\mu} \, \gamma^a
$
the local Dirac matrix, and consider its squared form:
\begin{equation}
\left(i \hbar \, \gamma^\mu \, {\mathcal D}_\mu - m c \right)
\left(i \hbar \, \gamma^\nu \, {\mathcal D}_\nu + m c \right) \psi = 0 \; .
\label{max1}
\end{equation}
By using the commutation relation
\begin{equation}
\left[{\mathcal D}_\mu, {\mathcal D}_\nu \right] = - \frac{i}{2} \,
{\stackrel{\circ}{R}} {}_{a b \mu \nu} \, \frac{\sigma^{a b}}{2} \; ,
\label{comd}
\end{equation}
with
$
{\stackrel{\circ}{R}}{}_{a b \mu \nu}
$
the curvature of the spin connection ${\stackrel{\circ}{\omega}}{}^{a}{}_{b \mu}$,
and
\begin{equation}
\sigma^{a b} = \frac{i}{2} \,
\left[\gamma^a, \gamma^b \right] \; ,
\label{s1/2re}
\end{equation}
we obtain
\begin{equation}
\left[- \, g^{\mu \nu} \, {\stackrel{\circ}{\mathcal D}}{}_{\mu} {\mathcal
D}{}_{\nu} + \frac{1}{2} \, \frac{\sigma^{\mu \nu}}{2} \,
{\stackrel{\circ}{R}}{}_{a b \mu \nu} \, \frac{\sigma^{a b}}{2} -
\frac{m^2 c^2}{\hbar^2} \right] \psi = 0 \; ,
\label{max2}
\end{equation}
where ${\stackrel{\circ}{\mathcal D}}_{\mu}$ stands for a Fock--Ivanenko
covariant derivative including a Levi--Civita connection
$\stackrel{\circ}{\Gamma}_{\mu}$ to account for the spacetime index appearing in the
first covariant derivative ${\mathcal D}{}_{\nu}$, and
\[
\sigma^{\mu
\nu} = h_{a}{}^{\mu} \, h_{b}{}^{\nu} \, \sigma^{a b} \; .
\]

Now, by using the identities ($i,j,k=1,2,3$) 
\[
\sigma^{0 k} \equiv i \alpha^{k} = i \,
\left( \begin{array}{cc} 0 & \sigma^k \\ \sigma^k & 0 \end{array} \right)
\; ; \; \sigma^{i j} = \epsilon_{ijk} \,
\left( \begin{array}{cc} \sigma^k & 0 \\ 0 & \sigma^k \end{array} \right) \; ,
\]
with $\sigma^k$ the Pauli spin matrices, and introducing the notation
\[
{B}_i = \frac{i}{2} \, \epsilon_{ijk} \, {\stackrel{\circ}{R}}_{j k 0 l} \, \alpha^l
\]
for the gravitomagnetic component of the curvature
tensor~\cite{wheeler2}, the curvature term of Eq.~(\ref{max2}) reads
\begin{equation}
\frac{1}{8} \, \sigma^{\mu \nu} \, {\stackrel{\circ}{R}}{}_{a b \mu \nu} \,
\sigma^{a b} = I \otimes \mbox{\boldmath$\mu$}_S
\mbox{\boldmath$\cdot B$} + \cdots \; ,
\end{equation}
where the dots denote terms corresponding to the other components of the curvature
tensor, and $I$ is a $2 \times 2$ unity matrix. Here
$\mbox{\boldmath$\mu$}_{S} = (2 / \, \hbar) \, \mbox{\boldmath$S$}$
is the adimensional gravitomagnetic moment of the electron,
with $\mbox{\boldmath$S$} = \hbar \mbox{\boldmath$\sigma$}/{2}$ the spin operator.
In units of gravitational Bohr magnetons, the gravitomagnetic moment is
\begin{equation}
\mbox{\boldmath$\mu$}_{S} = \frac{2 \; \mu_{B}}{\hbar} \, \mbox{\boldmath$S$}
\; .
\label{k=2}
\end{equation}
Comparing with Eq.~(\ref{mis}), we see that the quadratic form of the Dirac
equation in a Riemann spacetime, analogously to what occurs in the
electromagnetic case, yields the value $\kappa_{S} = 2$ for the spin
gravito--gyromagnetic ratio. We can, therefore, explicitly factorize the
gravitational gyromagnetic ratio $\kappa_S$, and rewrite Eq.~(\ref{max2}) in the form
\begin{equation}
\left[- \, g^{\mu \nu} \, {\stackrel{\circ}{\mathcal D}}{}_{\mu} {\mathcal
D}{}_{\nu} + \frac{\kappa_S}{4} \, \frac{\sigma^{\mu \nu}}{2} \,
{\stackrel{\circ}{R}}{}_{a b \mu \nu} \, \frac{\sigma^{a b}}{2} -
M^2 \right] \psi = 0 \; ,
\label{max25}
\end{equation}
where $M = m c/\hbar$.

\section{Field Equations}

In Minkowski spacetime, the square of the Dirac equation yields the
Klein--Gordon equation~\cite{bogo}, the field equation governing the dynamics
of a scalar field.
Behind this well known property, there is an implicit assumption according to which
all spinor representations of the Lorentz group must be replaced by scalar
representations. In a flat spacetime, however, this assumption is hidden by the fact
that those representations do not appear explicitly neither in the Dirac nor in the
resulting Klein--Gordon equation.

In a curved background, where the covariant derivatives are given by Fock--Ivanenko
operators, the situation is completely different. In fact, as we can see
from Eq.~(\ref{max2}), the representations of the Lorentz group appear explicit in
both the kinetic and the
curvature terms. Therefore, when looking for a similar result valid on a
curved background, such that in the limit of vanishing curvature the above Minkowski
space property be recovered, we have first to generalize Eq.~(\ref{max2}) by rewriting
it in an arbitrary representation $J^{a b}$ of the Lorentz group. Denoting by
$\Psi$ the corresponding arbitrary field, this means to write
\begin{equation}
\left[- \, g^{\mu \nu} \, {\stackrel{\circ}{\mathcal D}}{}_{\mu} {\mathcal
D}{}_{\nu} + \kappa_S \, \frac{\hbar^2}{4} \, J^{\mu \nu} \,  
{\stackrel{\circ}{R}}{}_{a b \mu \nu} \, J^{a b} - M^2 \right] \Psi = 0 \; .
\label{max3}
\end{equation}

Let us consider now the first few fundamental cases. When applied to a scalar field
$\phi$, the corresponding representation of the Lorentz group is $J^{a b} = 0$,
the Fock--Ivanenko operator becomes consequently an ordinary derivative,
and Eq.~(\ref{max3}) reduces to the Klein--Gordon equation in Riemann spacetime,
\begin{equation}
g^{\mu \nu} \, {\stackrel{\circ}{\nabla}}{}_{\mu} \, \partial_\nu \phi
+ M^2 \phi = 0 \; ,
\label{kg}
\end{equation}
We remark that the square--root of this
equation does {\it not} yield the Dirac equation (\ref{diraceq}), because some
information related to the coupling of the field spin with gravitation is lost
when the scalar condition $J^{a b} = 0$ is imposed. In curved spacetime, therefore,
the usual result stating that the Dirac equation is the square--root of the
Klein--Gordon equation does not hold. 

When applied to a Dirac spinor $\psi$, both $J^{\mu \nu}$ and $J^{a b}$ are replaced
by the spin--1/2 representation of the Lorentz group,
\[
J^{a b} = \frac{\sigma^{a b}}{2} \; ,
\]
and the gravitational gyromagnetic ratio assumes the value
$\kappa_S=2$. In this case, Eq.~(\ref{max3}) becomes
\begin{equation}
- g^{\mu \nu} \, {\stackrel{\circ}{\mathcal D}}{}_{\mu} {\mathcal D}{}_{\nu} \psi
+ \frac{1}{4} \, {\stackrel{\circ}{R}} \, \psi - M^2 \psi = 0 \; ,
\label{max4}
\end{equation}
which is the same as (\ref{max2}) with the second term reduced to its simplest form,
${\stackrel{\circ}{R}} = g^{\mu \nu}
{\stackrel{\circ}{R}}{}^{\rho}{}_{\mu \rho \nu}$
being the scalar curvature. This is the second--order
equation satisfied by a Dirac spinor.

When applied to a vector field $A^e$, supposed to satisfy the subsidiary condition
${\mathcal D}_\mu A^e = 0$, both $J^{\mu \nu}$ and $J^{a b}$
are replaced by the spin--1 representation of the Lorentz group,
\begin{equation}
\left(J^{a b} \right)^c{}_d \equiv \left(S^{a b} \right)^c{}_d =
i \left( \delta^{a}{}_{d} \, \eta^{b c} - \delta^{b}{}_{d} \, \eta^{a c}
\right) \; .
\label{vecre}
\end{equation}
In this case, Eq.~(\ref{max3}) becomes
\begin{equation}
\left( {\stackrel{\circ}{\nabla}}{}_{\mu} \, {\stackrel{\circ}{\nabla}}{}^{\mu}
- m^2 \right) A_\nu - \kappa_S \, {\stackrel{\circ}{R}}{}^{\mu}{}_{\nu} \, A_\mu = 0 \; ,
\label{max}
\end{equation}
and the subsidiary condition acquires the form
\[
{\stackrel{\circ}{\nabla}}{}_{\mu} A^\mu = 0 \; .
\]
If we assume the
gravitational gyromagnetic ratio for a vector field to be
$\kappa_S = 1$, the above equation turns out to be Proca's equation in a
Riemann spacetime.

\section{Final Remarks}

Equation (\ref{max3}) is the central second--order equation for fields of spin 0, 1/2
and 1, provided the correct value for $\kappa_S$ is inserted in each
case. In particular, a gravitational gyromagnetic ratio
$\kappa_S = 2$ is found for the interaction of the spin of a Dirac spinor with
curvature. A gravitational gyromagnetic ratio $\kappa_S = 1$ has also been obtained 
for a vector field. Analogously to what occurs
in Min\-kows\-ki spacetime, the Dirac equation in a Riemann spacetime can be
obtained by means of a factorization of the corresponding second--order Proca's equation,
but this is possible only if the gravitational gyromagnetic ratios are
properly taken into account. For example, taking the square--root of the Riemannian
Proca operator appearing in (\ref{max}) does yield the Dirac
equation~(\ref{diraceq}), provided in the process of taking the square--root the
gravitational gyromagnetic ratio $\kappa_S$ is properly set equal to $2$, which is
the correct value for a Dirac field. It is important to remark, however, that the Dirac
equation cannot be obtained as the square--root of the Klein--Gordon operator, as
some  essential information concerning the spin--gravitation coupling is lost
when the scalar condition $J^{a b} = 0$ is imposed. Therefore, we conclude that,
by taking into account the correct value of the gravitational gyromagnetic ratios,
it turns out possible to construct a general Laplacian~\cite{rose} valid for both
integer and half--integer spin fields,
unifying in this way the usual treatment of these fields~\cite{chris}.

The experimental detection of gravitomagnetic effects is still at its
beginnings~\cite{probe}. At the macroscopic level, experiences involving orbiting
gyroscopes to probe Earth's gravitomagnetism have just yielded their first
results~\cite{ciufolini}. At the microscopic, elementary particle level no
experimental results are expected in the near future. Our results are, however,
essential to the self--consistency of gravitation theory and its coherence with
field theory in general.

\section*{Acknowledgments}

The authors would like to thank I. B. Khriplovich for useful comments, and for
calling our attention to the paper of Ref.~\cite{khri}, where a similar study has
been performed. They would like also to thank G. F. Rubilar for useful discussions.
This work was supported by FAPESP--Brazil and CNPq--Brazil.

\end{document}